\documentstyle{elsartwb}

\input epsf
\input psfig

\begin{document}

\begin{frontmatter}

\title{$\pi^0 \to \gamma \gamma^\ast$,
$\omega \to \pi^0  \gamma^\ast$,
and $\rho \to \pi \gamma^\ast $
decays in nuclear medium\thanksref{grants}}
\thanks[grants]{Research supported by
        the Polish State Committee for
        Scientific Research grant 2P03B09419.}

\author{Agnieszka Bieniek, Anna Baran, Wojciech Broniowski}

\address{The H. Niewodnicza\'nski Institute of
Nuclear Physics, PL-31342 Krak\'ow, Poland}

\begin{abstract}
We calculate the medium modification of 
the $\pi \omega \rho$ vertex and analyze its
significance for the $\pi^0 \to \gamma \gamma^\ast$,
$\omega \to \pi^0  \gamma^\ast$,
and $\rho \to \pi \gamma^\ast$ decays in nuclear matter.
We use a relativistic hadronic
approach at zero temperature and in the leading-density
limit, and consider decays of particles at rest with respect to the medium.
It is shown that for the  $\pi^0 \to \gamma \gamma^\ast$ the
effects of the $\Delta$ isobar cancel almost exactly the effects
of the nucleon $ph$ excitations, such that the net medium
effect is small. On the contrary, for the decays
$\omega \to \pi^0 \gamma^\ast$ and $\rho \to \pi \gamma^\ast$ we find a
sizeable increase of the partial widths at virtualities of the photon
in the range $0.3-0.6$GeV. The effect has direct significance for the
calculation of dilepton yields from the Dalitz decays in relativistic
heavy-ion collisions.
\end{abstract}


\end{frontmatter}

\vspace{-5mm} PACS: 25.75.Dw, 21.65.+f, 14.40.-n

In recent years a lot of efforts have been undertaken in order to understand
and quantitatively describe the properties of hadrons in nuclear matter. As
an outcome, it has been commonly accepted that hadrons undergo substantial
modifications induced by the medium \cite
{brscale,celenza,serot,chin,hatsuda,jean,herrmann1,%
pirner,rapp,LeupoldRev,Gao,hatlee,lee2000,%
Vogl,klingl,Chanfray,BH1,eletsky,FrimanActa,Lutz99},
a view supported by numerous studies of two-point functions. Since the
masses and widths of hadrons are significantly modified, one expects that
also coupling constants are altered. However, there exist only a few
investigations devoted to the issue of in-medium hadronic {\em three-point
functions}
\cite{Song2,BFH1,BFH2,Urban0,urban,Banerjee96,rhopipi}. 
In particular, it
has been found that the $\rho \pi \pi $ coupling is largely increased in
nuclear matter. This suggests the necessity of a careful examination of
other couplings as well. In this paper we examine the {\em in-medium}
$\pi \omega \rho$ vertex, which appears in such processes as $\pi
^{0}\rightarrow \gamma \gamma ^{\ast }$, $\omega \rightarrow \pi ^{0}\gamma
^{\ast }$, and $\rho \rightarrow \pi \gamma ^{\ast }$. These processes
are important for theoretical studies of the dilepton production via Dalitz
decays in relativistic heavy-ion collisions. To our knowledge, the existing
calculations and numerical codes use the vacuum value of the $\pi \omega
\rho $ vertex. Clearly, medium effects, if present, alter the
predictions of such calculations and this is the reason of our study.
We find that for the Dalitz decays of vector mesons the 
presence of nuclear matter leads to a sizeable increase of partial widths:
a factor of 2 for the $\omega$ and a factor of 3 for the $\rho$ at
the photon virtualities around 0.4GeV.
This increase, in turn, directly leads to enhanced dilepton yields
in processes $\omega \rightarrow \pi ^{0}\gamma l \overline{l}$
and $\rho \rightarrow \pi \gamma l \overline{l}$ , which may help to
resolve the long-standing problem of the low-mass dilepton
enhancement seen in relativistic heavy-ion collision \cite{ceres,helios}.

In our analysis we apply a conventional hadronic model with meson, nucleon
and $\Delta$-isobar degrees of freedom, and work at the leading-density
and zero temperature. We constrain ourselves to the situation where the
decaying particle is at rest with respect to the medium. The extension to the
general kinematic case is straightforward and will be presented elsewhere,
together with the non-zero temperature analysis.

The vacuum value of the $\pi \omega \rho $ vertex (Feynman rule) is
\begin{equation}
-iV_{\pi \omega \rho }^{\nu \mu }=i\frac{g_{\pi \omega \rho }}{F_{\pi }}%
\epsilon ^{\nu \mu pQ},  \label{vac}
\end{equation}
where we have used the convenient short-hand notation $\epsilon ^{\nu \mu
pQ}=\epsilon ^{\nu \mu \alpha \beta }p_{\alpha }Q_{\beta }$. We choose $Q$ as
the incoming momentum of the $\pi $, and $p$ as the outgoing momentum of the
$\rho $, and $q\equiv Q-p$ is the outgoing momentum of the $\omega $. In a
general case $\omega $ and $\rho $ are a virtual particles. The value of $%
g_{\pi \omega \rho }$ is well known. For the case where all external momenta
vanish, it can be obtained with help of the vector meson dominance from the
anomalous $\pi \gamma \gamma $ coupling constant, namely $g_{\pi \omega \rho
}=-\frac{g_{\rho }g_{\omega }}{e^{2}}g_{\pi \gamma \gamma }$, with $g_{\pi
\gamma \gamma }=\frac{e^{2}}{4\pi ^{2}F_\pi}$ fixed by the anomaly.

Our calculation is made in the framework of a relativistic hadronic theory,
where mesons interact with the nucleons and $\Delta (1232)$ isobars.
The in-medium nucleon propagator can be conveniently decomposed into
the {\em free} and {\em density} parts \cite{chin}:
\begin{eqnarray}
S(k) &=&S_{F}(k)+S_{D}(k)=i(\gamma ^{\mu }k_{\mu }+m_{N})[\frac{1}{%
k^{2}-m_{N}^{2}+i\varepsilon }+  \nonumber \\
&&\frac{i\pi }{E_{k}}\delta (k_{0}-E_{k})\theta (k_{F}-|k|)],  \label{S}
\end{eqnarray}
where $m_{N}$ denotes the nucleon mass, $E_{k}=\sqrt{m_{N}^{2}+k^{2}}$, and $%
k_{F}$ is the Fermi momentum.
The Rarita-Schwinger $\Delta $ propagator \cite{rarita,Ben} has the form
\begin{equation}
iS_{\Delta }^{\mu \nu }(k)=i\frac{\gamma ^{\mu }k_{\mu }+M_{\Delta }}{%
k^{2}-M_{\Delta }^{2}}(-g^{\mu \nu }+\frac{1}{3}\gamma ^{\mu }\gamma ^{\nu }+%
\frac{2k^{\mu }k^{\nu }}{3M_{\Delta }^{2}}+\frac{\gamma ^{\mu }k^{\nu
}-\gamma ^{\nu }k^{\mu }}{3M_{\Delta }}). \label{rarita}
\end{equation}
We incorporate phenomenologically the effects of the finite width of the $%
\Delta $ by the replacement $M_{\Delta }\rightarrow M_{\Delta }-i\Gamma
/2 $.\footnote{This treatment of the finite width of the $\Delta$
is consistent with the Ward-Takahashi identities for the $\pi \rho \omega$
vertex. This would not be true if $\Gamma_\Delta$ were 
introduced in the denominatorof Eq. (\ref{rarita}) only.}

\begin{figure}[t]
\centerline{\psfig
{figure=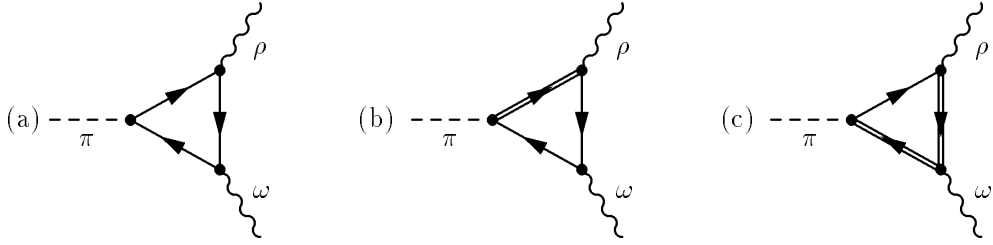,height=5cm,bbllx=75bp,bblly=377bp,bburx=541bp,bbury=537bp,clip=}}
\label{diag}
\caption{Diagrams included in our calculation (crossed diagrams not
shown). Wavy lines indicates the $\protect\rho $, dashed lines the pions,
solid lines the in-medium nucleon, and double lines the $\Delta $.}
\end{figure}

The diagrams included in the calculation are displayed in Fig. 1, with the
wavy lines denoting the $\rho $ or $\omega $, the dashed line the $\pi ^{0}$%
, the double line the $\Delta $, and the single line the in-medium nucleon.
Fermi-sea effects arise when one of the nucleon lines in each of the our
diagrams of Fig. 1 involves the nucleon density propagator, $S_{D}$ . For
kinematic reasons, diagrams with more than one $S_{D}$ vanish. Diagrams with
no $S_{D}$ propagator do not describe Fermi-sea effects; these
vacuum-polarization diagrams are not considered in the present study.

The vertices (Feynman rules) needed for our calculation have the form
\begin{eqnarray}
-iV_{\omega NN} &=&-ig_{\omega }\gamma ^{\mu }, \\
-iV_{\rho NN} &=&i\frac{g_{\rho }}{2}(\gamma ^{\mu }-\frac{i\kappa _{\rho }}{%
2m_{N}}\sigma ^{\mu \nu }p_{\nu })\tau ^{b}, \\
-iV_{\pi NN} &=&\frac{g_{A}}{2F_{\pi }}\FMslash{Q}\gamma ^{5}\tau ^{a},
\end{eqnarray}
where $p$ is the outgoing momentum of the virtual $\rho $ , $Q$ is incoming
four-momentum of the $\pi $, and $a$, and $b$ are the isospin indices of the
$\pi $, and $\rho $ respectively. For the interactions involving the $\Delta $
we have \cite{Ben}
\begin{eqnarray}
-iV_{N\Delta \pi } &=&g_{N\Delta \pi }Q^{\alpha }T^{a}, \\
-iV_{N\Delta \rho } &=&ig_{N\Delta \rho }(\FMslash{p}\gamma ^{5}g^{\alpha
\beta }-\gamma ^{\beta }\gamma ^{5}p^{\alpha })T^{b}, \\
-iV_{\Delta \Delta \omega } &=&-ig_{\omega }(\gamma ^{\alpha }\gamma ^{\beta
}\gamma ^{\delta }-\gamma ^{\delta }g^{\alpha \beta }+\gamma ^{\alpha
}g^{\beta \delta }+\gamma ^{\beta }g^{\delta \alpha }),
\end{eqnarray}
where $T^{a}$ and $T^{b}$ are the standard isospin $\frac{1}{2}\rightarrow
\frac{3}{2}$ transition matrices \cite{EW}. Our choice of physical
parameters is as follows:
\begin{eqnarray}
g_{\omega } &=&10.4,\quad g_{\rho }=5.2,\quad \quad g_{A}=1.26,\quad F_{\pi
}=93{\rm MeV},  \nonumber \\
g_{N\Delta \pi } &=&\frac{2.12}{m_{\pi }},\quad g_{N\Delta \rho }=\frac{2.12%
\sqrt{2}}{m_{\pi }}.  \label{par}
\end{eqnarray}
The values of the $N\Delta $ coupling constants follow from the
non-relativistic reduction of the vertices and comparison to the
non-relativistic values \cite{EW}. The $\Delta \Delta \omega $ coupling
incorporates the principle of universal coupling. In the results presented
below we have chosen 
$\kappa _{\rho }=3.7$,{\em \ i.e.} $\kappa _{\rho }=\kappa _{V}
$ value. Qualitatively similar conclusions follow when $\kappa _{\rho }=6$
is used. One should note here that the relativistic form of couplings to the
$\Delta $, as well as the values of the coupling constants, are subject of
an on-going discussion and research \cite{Ben,Pascal,Hemmert,Haber}.
However, the ambiguities involved 
are not relevant for our study, where we explore
the possibility and the size of the effect rather than seek accurate
predictions.

In the following the results of the medium-modified vertex at the
nuclear-saturation density ($\rho _{0}=0.17{\rm fm}^{-3}$) will be presented
relative to the vacuum value $g_{\rm vac}=g_{\pi \rho \omega}$. We define
\begin{equation}
g_{\rm eff}=g_{\pi \rho \omega }+\rho _{0}B,  \label{geff}
\end{equation}
with $B$ denoting the medium contribution evaluated according to the
diagrams of Fig. 1. The in-medium $\pi \omega \rho $ vertex has the general
tensor structure
\begin{eqnarray}
A^{\mu \nu } &=&A_{1}\varepsilon ^{\mu \nu pQ}+A_{2}\varepsilon ^{\mu \nu
uQ}+A_{3}\varepsilon ^{\mu \nu pu}+A_{4}\varepsilon ^{\mu upQ}p^{\nu }
\label{struct} \\
&&+A_{5}\varepsilon ^{\mu upQ}Q+A_{6}\varepsilon ^{\mu upQ}u^{\nu
}+A_{7}\varepsilon ^{\nu upQ}p^{\mu }+A_{8}\varepsilon ^{\nu upQ}Q^{\mu
}+A_{9}\varepsilon ^{\nu upQ}u^{\mu }.  \nonumber
\end{eqnarray}
This structure, restricted by the Lorentz invariance and parity, is more
general than in the vacuum case (\ref{vac}) due to the 
availability of the
four-velocity of the medium, $u$. The result of {\em any} 
calculation will assume
the form (\ref{struct}). The leading-density calculation of diagrams of Fig.
1 is very simple. As has been shown {\em e.g.} in Ref. \cite{rhopipi}, it amounts to
evaluating the trace factors, substituting $k^{\mu }\rightarrow m_{N}u^{\mu }
$, with $u^{\mu }=(1,0,0,0)$ in the rest frame of the medium, and replacing
the loop momentum integration by $\int \frac{d^{4}k}{(2\pi )^{4}}\Theta
(k_{f}-\left| \vec{k}\right| )\delta (k_{0}-E_{k})\rightarrow \frac{1}{8\pi }%
\rho _{B},$ where $\rho _{B}$ is the baryon density. The vertex $A^{\mu \nu }
$ satisfies the Ward-Takahashi identities: $q_{\mu }A^{\mu \nu }=0,$ and $%
p_{\nu }A^{\mu \nu }=0$, which serves as a useful check of the algebra.

We begin the presentation of our numerical results with the process $\pi
^{0}\rightarrow \gamma \gamma $ (decay into two real photons), where the
pion is {\em at rest} with respect to the medium. In this case the four-vectors $Q$
and $u$ are parallel, and out of the $9$ structures in Eq. (\ref{struct})
only $A_{1}$ and $A_{3}$ survive: $A^{\mu \nu }=A_{1}\varepsilon ^{\mu \nu
pQ}+A_{3}\varepsilon ^{\mu \nu pu}$. Thus in the rest frame of the medium,
where $Q=(m_{\pi },0,0,0),$ we find $A^{\mu \nu }=A_{1}Q_{0}\varepsilon
^{\mu \nu p0}+A_{3}u_{0}\varepsilon ^{\mu \nu p0}=(m_{\pi
}A_{1}+A_{3})\varepsilon ^{\mu \nu p0}=(A_{1}+A_{3}/m_{\pi })\varepsilon
^{\mu \nu pQ}$. We combine the vacuum and medium pieces and 
introduce the notation $A^{\mu \nu }=-i\frac{%
e^{2}}{g_{\rho }g_{\omega }}\left( \frac{g_{\pi \rho \omega }}{F_{\pi }}-%
\frac{\rho _{B}}{F_{\pi }}B\right) \epsilon ^{\nu \mu pQ}$, which places the
medium effects in the constant $B$. The contributions to $B$ from diagrams
(a), (b) and (c) of Fig.1 are as follows:
\begin{eqnarray}
B_{(a)} &=&\frac{g_{A}g_{\rho }g_{\omega }(\kappa +1)}{m_{N}(4m_{N}^{2}-m_{%
\pi }^{2})}, \\
B_{(b)} &=&\frac{g_{N\Delta \pi }g_{N\Delta \rho }g_{\omega }F_{\pi }m_{\pi }%
}{9M_{\Delta }^{2}m_{N}[(m_{N}^{2}-M_{\Delta }^{2})^{2}-m_{\pi
}^{2}m_{N}^{2}]}(m_{N}^{4}+M_{\Delta }m_{N}^{3}-2M_{\Delta }^{2}m_{N}^{2}- \\
&&-m_{\pi }^{2}m_{N}^{2}-M_{\Delta }^{3}m_{N}+M_{\Delta }^{4}),  \nonumber \\
B_{(c)} &=&\frac{g_{N\Delta \pi }g_{N\Delta \rho }g_{\omega }F_{\pi }m_{\pi }%
}{27M_{\Delta }^{4}[(m_{N}^{2}-M_{\Delta }^{2})^{2}-m_{\pi }^{2}m_{N}^{2}]}%
(-4m_{N}^{5}-6M_{\Delta }m_{N}^{4}+8M_{\Delta }^{2}m_{N}^{3}+ \\
&&+5M_{\Delta }^{3}m_{N}^{2}+(4m_{N}+3M_{\Delta })m_{\pi
}^{2}m_{N}^{2}-16M_{\Delta }^{4}m_{N}-11M_{\Delta }^{5}),  \nonumber
\end{eqnarray}
where $M_{\Delta }$ is understood to carry the width 
$i \Gamma _{\Delta }$. With
parameters (\ref{par}) we find numerically $B_{(a)}=97{\rm GeV}^{-3}$, $%
B_{(b)}=-3.1{\rm GeV}^{-3}$, $B_{(c)}=-102{\rm GeV}^{-3}$ for the formal
case $\Gamma _{\Delta }=0$. For the physical vacuum value of the $\Delta $
width, $\Gamma _{\Delta }=0.12{\rm GeV},$ we find $B_{(a)}=97{\rm GeV}^{-3}$%
, $B_{(b)}=-(2.6+1.5i){\rm GeV}^{-3}$, $B_{(c)}=-(87+42i){\rm GeV}^{-3}$ and
$|g_{\rm eff}/g_{\rm vac}|^{2}=\left| 0.99+0.04i\right| ^{2}=0\,.99,$ whereas with
nucleons only (diagram (a)) we would have $%
|g_{\rm eff}^{(a)}/g_{\rm vac}|^{2}=0.91^{2}=0.82$. Thus
the effects of the $\Delta $ act in the opposite direction than the
nucleons, resulting in an almost complete cancellation between diagrams (a)
and (b,c) of Fig. 1.

The Dalitz decay of $\pi ^{0}$ into the photon and the lepton pair
proceeds through the decay into a real and a virtual photon, and subsequently
the decay of the virtual photon into the lepton pair,
$\pi ^{0}\rightarrow \gamma \gamma ^{\ast }
\rightarrow \gamma l \overline{l}$.
The virtual photon can be either isovector ($%
\rho $-type) or isoscalar ($\omega$-type).
We denote the virtuality of $\gamma ^{\ast
}$ as $M$, and investigate the dependence of $g_{\rm eff}$ on $M$. The results,
displayed in Fig. 2, show that the change of 
$M$ in the allowable kinematic
range from $0$ to $m_{\pi }$ has a very little effect on the ratio of
$|g_{\rm eff}/g_{\rm vac}|^{2}$. We notice a somewhat different behavior for the
isoscalar and isovector photons at $M$ approaching $m_{\pi }$, with the
former ones coupled with greater strength. By comparing the curves
corresponding to the
calculation with the $\Delta $ (diagrams (a+b+c) of Fig. 1) and with
nucleons only (diagrams (a)), we can see that the
cancellation between diagrams (a) and (b+c) hold for the 
whole kinematically-available range. As a result, the medium effects on
the Dalitz decays of $\pi ^{0}$ are negligeable, at most at the level of 
a few percent.

\begin{figure}[tbp]
\vspace{0mm} \epsfxsize = 11 cm \centerline{\epsfbox{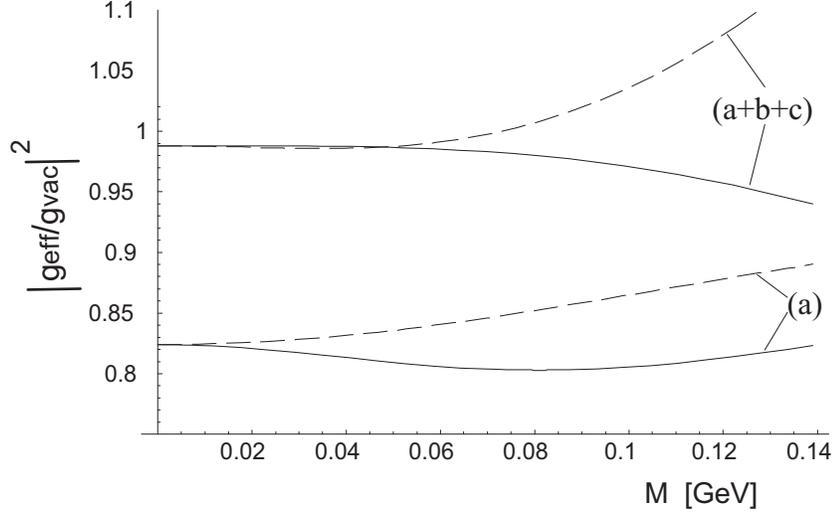}} \vspace{0mm}
\label{fig:diag}
\caption{The quantity $|g_{\rm eff}/g_{\rm vac}|^{2}$ for the process $%
\pi ^{0}\rightarrow \gamma ^{\ast }\gamma ,$ plotted as a function of the
virtual mass of $\gamma ^{\ast }$, $M$, for the case $\rho _{B}=\rho _{0}$
and  $\Gamma _{\Delta }=120{\rm MeV}$. The solid and dashed lines
corresond to the case of isovector and isoscalar $\gamma^\ast$,
respectively. The upper curves show the results of the full
calculation, with diagrams (a,b,c) of Fig. 1 included, while the lower
curves show the result of the calculation with diagram (a) only.}
\end{figure}

Next, we ``turn around'' the diagrams of Fig. 1 in order to consider the
processes $\omega \rightarrow \pi ^{0}\gamma ^{\ast }$ and $\rho \rightarrow
\pi ^{0}\gamma ^{\ast }$, where $\omega $ and $\rho$ are on mass shell.
In these cases we use the vacuum value $g_{\rho \omega \pi}=-1.13$,
inferred from
the experimental partial decay width of
$\omega \to \pi^0 \gamma$, rather than
the value  $g_{\pi \omega \rho
}=-\frac{g_{\rho }g_{\omega }}{e^{2}}g_{\pi \gamma \gamma }=-1.36$
used earlier for the $\pi^0$ decay. The difference may be 
attributed to a form-factor effect:
the virtuality of the vector meson is changed from 0 to the on-shell mass
of the vector meson.

There is an interesting phenomenon in the decays of vector mesons,
related to the analyticity of the
amplitudes of Fig. 1 (b,c) in the virtuality of $\gamma^{\ast}$, denoted as
$M$. For the case $\Gamma_{\Delta }=0$ the amplitudes develop a pole at
the location
\begin{equation}
M=\sqrt{\frac{m_{v}m_{N}^{2}+m_{\pi }^{2}m_{N}+m_{v}^{2}m_{N}-M_{\Delta
}^{2}m_{v}+m_{\pi }^{2}m_{v}}{m_{N}}}\approx 0.34{\rm GeV},
\end{equation}
where $m_{v}=m_{\rho}$ or $m_{\omega}$.
The pole can be clearly seen in the plots of Fig. 3 for the case of
$\Gamma_\Delta=0$ (dashed lines). For the vacuum 
value of $\Gamma_\Delta-120$MeV
(solid lines) the pole is washed-out, but its traces are still visible, with
the curves reaching maxima around $M=0.45$GeV. We
observe that at low values of $M$ the effective coupling constant
remains unchanged for the $\omega$ decay, and  is decreased for the $\rho$
decay. However, at higher values of $M$, above 0.2GeV, the value
of $g_{\rm eff}$ is larger
than in the vacuum, with the effect quite large: an enhancement
by a factor of 2
for the $\omega$ decay, and by a factor of 3 for the $\rho$ decay.
The exacts numbers have to treated with a grain of salt, since we have assumed
the low-density limit, good as long as the results remain small. However, 
the values obtained are strongly indicative of possible large effects 
in the Dalitz decays of vector mesons in medium.

\begin{figure}[tb]
\vspace{0mm} \epsfxsize = 9 cm \centerline{\epsfbox{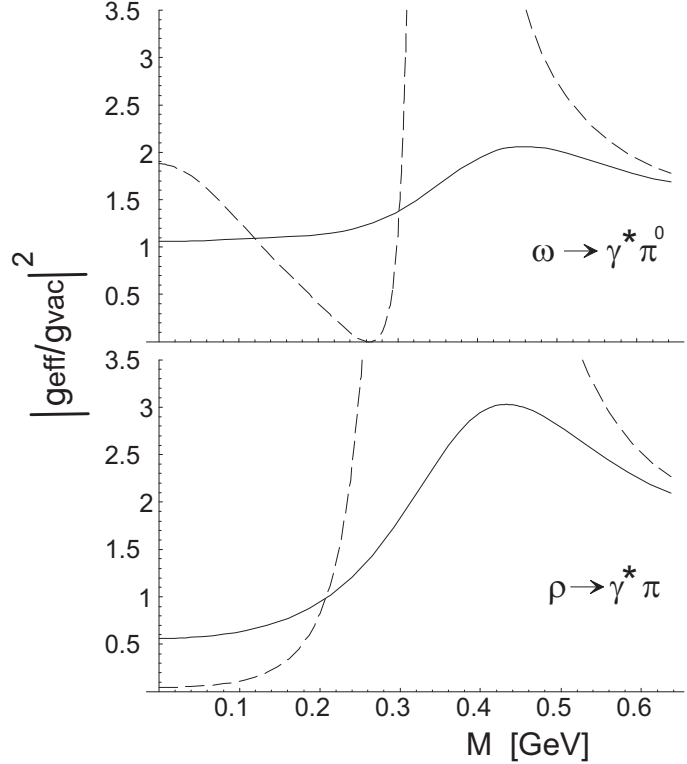}}
\vspace{0mm} \label{fig:3}
\caption{Top: $|g_{\rm eff}/g_{\rm vac}|^{2}$ for the
decay $\omega \to \gamma^\ast \pi^0$ at the nuclear saturation
density, plotted as a function of the
virtuality of the photon, $M$.
The dashed and solid lines correspond to the cases
$\Gamma_\Delta=0$ and $120{\rm MeV}$, respectively. Bottom:
the same for the decay $\rho \to \gamma^\ast \pi$.}
\end{figure}

The enhancement of Fig. 3 
will directly affect the calculations of the dilepton
production in relativistic heavy-ion collisions, which is in fact very
much desired. According to the results presented above, 
the theoretical yields from the
Dalitz decays of vector mesons  become considerably enhanced
(a factor of 2 for the dominating $\omega$ meson) 
in the region of $0.4-0.5$GeV, which is precisely where many existing
calculations have problems in providing enough dileptons 
to explain the excess seen in the CERES and HELIOS experiments.
At the same time, the Dalitz yields from the $\pi^0$ decays are not altered, 
which is also desired, as the theoretical calculations with the vacuum
value of the $\pi \gamma \gamma^\ast$ vertex explain 
accurately the dilepton data in the low-mass (up to 0.2GeV) region.

The authors are grateful to Wojciech Florkowski and Antoni Szczurek for
numerous useful conversations. 

\bibliographystyle{npsty}
\bibliography{delta,wb}

\end{document}